\documentclass[journal]{IEEEtran}
\ifCLASSINFOpdf
\usepackage[pdftex]{graphicx}
\graphicspath{figures/}
\DeclareGraphicsExtensions{.pdf,.jpeg,.png}
\else
\usepackage[dvips]{graphicx}
\graphicspath{{figures/}}
\DeclareGraphicsExtensions{.eps}
\fi
\usepackage[cmex10]{amsmath}
\usepackage{bm}
\usepackage{hyperref}
\usepackage{mathtools}

\usepackage{url}
\begin{document}
\title{Probabilistic Deep Spiking Neural Systems Enabled by Magnetic Tunnel Junction}

\author{Abhronil~Sengupta,~\IEEEmembership{Student Member,~IEEE,}
Maryam~Parsa, Bing~Han,
and~Kaushik~Roy,~\IEEEmembership{Fellow,~IEEE}
\thanks{The work was supported in part by, Center for Spintronic Materials, Interfaces, and Novel Architectures (C-SPIN), a MARCO and DARPA sponsored StarNet center, by the Semiconductor Research Corporation, the National Science Foundation, Intel Corporation and by the National Security Science and Engineering Faculty Fellowship.}
\thanks{The authors are with the School
of Electrical and Computer Engineering, Purdue University, West Lafayette,
IN, 47907 USA. E-mail: asengup@purdue.edu.}}
\maketitle
\begin{abstract}
\small{Deep Spiking Neural Networks are becoming increasingly powerful tools for cognitive computing platforms. However, most of the existing literature on such computing models are developed with limited insights on the underlying hardware implementation, resulting in area and power expensive designs. Although several neuromimetic devices emulating neural operations have been proposed recently, their functionality has been limited to very simple neural models that may prove to be inefficient at complex recognition tasks. In this work, we venture into the relatively unexplored area of utilizing the inherent device stochasticity of such neuromimetic devices to model complex neural functionalities in a probabilistic framework in the time domain. We consider the implementation of a Deep Spiking Neural Network capable of performing high accuracy and low latency classification tasks where the neural computing unit is enabled by the stochastic switching behavior of a Magnetic Tunnel Junction. Simulation studies indicate an energy improvement of $20\times$ over a baseline CMOS design in $45nm$ technology. }
\end{abstract}


\section{Introduction}
Despite the huge success of Deep Artificial Neural Networks (ANN) at complex recognition problems like the CIFAR \cite{krizhevsky2009learning} and ImageNet \cite{deng2009imagenet} benchmarks, the significant computational costs involved in training and testing such deep nets have inspired researchers to develop alternative computing models. Over the past few years, ANNs have evolved into the more biologically realistic Spiking Neural Nets (SNN) where information is communicated between the neural nodes as spikes rather than real-valued analog signals. Such spiking networks have resulted in the development of specialized custom hardware implementations \cite{merolla2014million} that exploit the prospects of event-based computing. However, training such SNNs for recognition problems have been mostly limited to single-layered networks \cite{diehl2015unsupervised} which have been unable to compete with the high recognition performances offered by deep ANN networks. Hence, research efforts have been directed to develop algorithms for converting a fully trained ANN computing model to a corresponding SNN model in order to achieve event-driven hardware implementation \cite{diehl2015fast,hunsberger2015spiking}. However, such conversion schemes have been developed with little or no regard to the underlying hardware implementation for the neuron or synaptic units. 

As a parallel effort, research in neuromorphic computing has been aimed at identifying nanoelectronic devices that can mimic and thereby offer a compact and energy-efficient implementation for neural and synaptic units. While much research has been conducted on the implementation of synaptic functionalities like Spike-Timing Dependent Plasticity \cite{jackson2013nanoscale,kuzum2011nanoelectronic,jo2010nanoscale,sengupta2015spin} and Short-Term Plasticity effects \cite{ohno2011short,chang2011short,PhysRevApplied.5.024012} in resistive technologies, potential ``neuristor'' devices emulating neuronal units are still in its infancy. For instance, spintronic devices have been proposed to be a promising candidate for implementing such neural functionalities \cite{sharad2012spin,sharad2013spin,sengupta2016spin} but have been able to implement only the ``step'' (neuron switching state depending on sign of input stimulus) transfer function of ANNs while a graded analog ANN transfer function like the sigmoid can be potentially appealing for implementing deep ANNs capable of performing complex recognition tasks. It is worth noting here that although recent proposals have investigated the implementation of analog ANN transfer functions by spintronic devices \cite{fan2015stt}, they require fabrication of relatively complex device structures based on multi-domain nanomagnets.

This work lies at the juncture of the two parallel research thrusts mentioned above. We note that although technologically mature spintronic devices, like the Magnetic Tunnel Junction (MTJ) based on mono-domain magnets, may not be able to exhibit complex analog ANN neural transfer functions (being binary switching devices), they exhibit switching probability characteristics that vary in a fashion similar to the sigmoid function with variation in the magnitude of the input current. Based on this observation, we propose an ANN to SNN conversion scheme by arguing that a fully trained ANN can be converted to an SNN if the neural units are assumed to generate spikes depending on a probability density function which is similar to the original ANN transfer function. We provide a mathematical formulation to justify that such a conversion mechanism is able to approximate the original ANN functionality to a reasonable degree of precision. Our motivation is driven by the fact that in addition to being an intuitive formulation for ANN-SNN conversion, such an implementation can be enabled by the underlying stochastic device physics of the MTJ. This proposal can potentially pave the way for probabilistic neuromorphic platforms that exploit the device variability and stochasticity inherent in such emerging neuromimetic devices.

\section{Probabilistic Spiking Neural Computation: Conversion from ANN}
In this section, we will describe the neural computational unit typically used for ANNs, followed by an intuitive discussion for the basis of our proposed conversion mechanism to SNN. Subsequently, we will illustrate a simple mathematical justification to validate our claims. 
\begin{figure}[!t]
\centering
\includegraphics[width=3.4in]{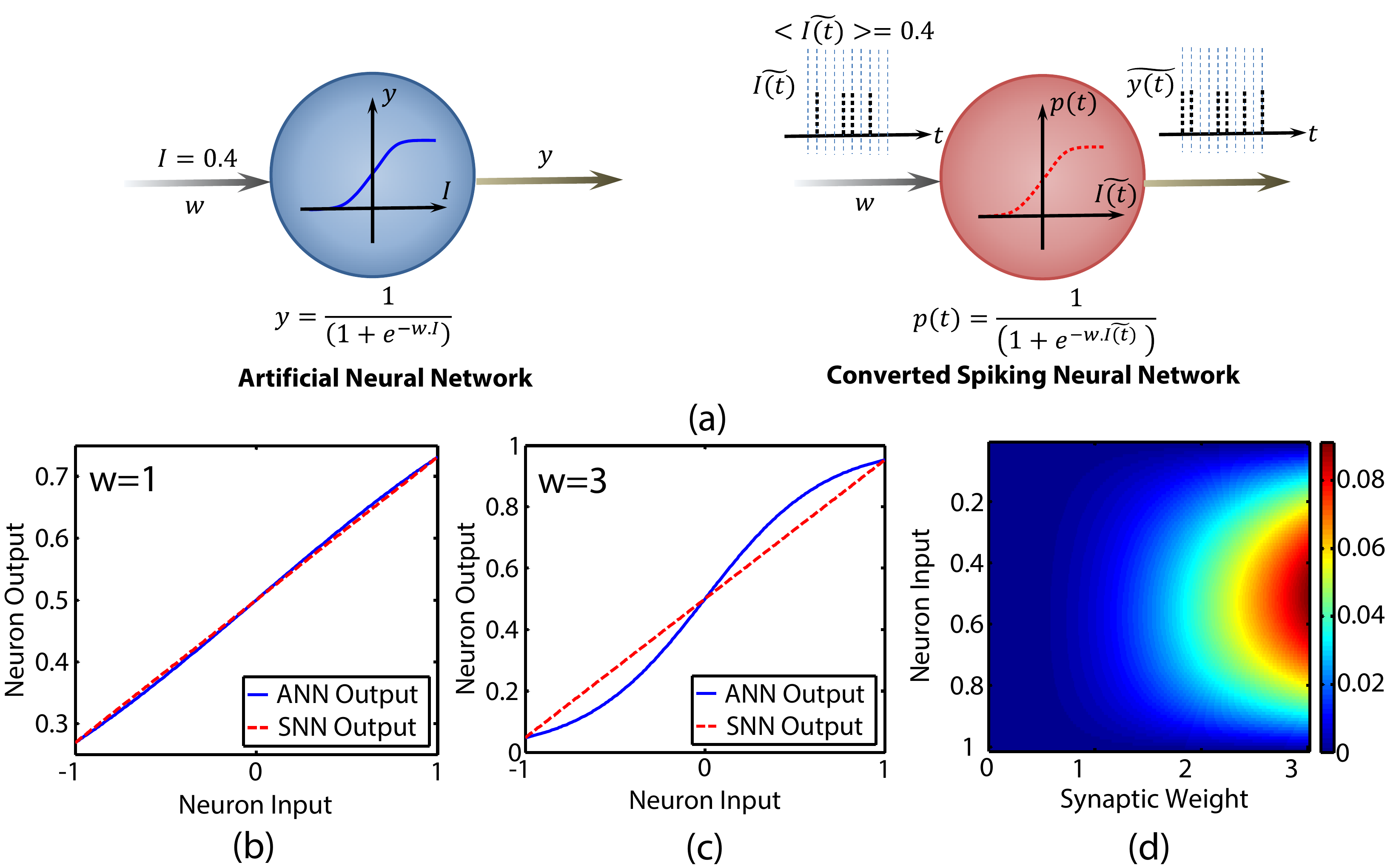}
\caption{(a) The ANN is converted to SNN computing model by interpreting the neuron transfer function as the neuron spiking probability in the SNN mode, (b) and (c) ANN and SNN outputs are plotted over the entire input range for weight magnitudes, $w=1$ and $w=3$ (maximum weight) respectively, (c) Error contour plot between the ANN output and the converted SNN output with variation in both neuron input and synaptic weight magnitudes. The error increases with increasing weight but remains bounded within reasonably low values.}
\label{fig1}
\end{figure}

Let us consider an ANN neural unit that receives an input $I$ through a synapse of weight $w$. The neuron generates an output $y$ by passing the weighted input through a non-linearity $f(.)$. We will consider the function $f(.)$ to be the sigmoid function $\left(f(x)=\frac{1}{1+e^{-x}}\right)$ in this work, due to its popularity in traditional ANN networks for achieving high accuracy in complex recognition problems \cite{IMM2012-06284} along with the possibility of enabling this functionality by MTJ devices, as will be explained in the next section. Hence, for the ANN neuron, the corresponding output $y$ will be given by,
\begin{equation}
y=\frac{1}{1+e^{-w.I}}
\end{equation}
It is worth noting here that the input $I\in[0,1]$, since it represents the inputs coming from normalized values of external stimuli (image pixels for image recognition systems) or from other neuron outputs in previous layers (which lie in the range $[0,1]$ due to the limited range of sigmoid function). 

Next, let us describe the proposed conversion process from ANN to SNN (Fig. \ref{fig1}(a)). In the spiking mode of communication, the input $I$ can be rate encoded as a Poisson spike train $\widetilde{I}(t)$. The train consists of a sufficiently large number of time-steps, $T_{N}$, where the probability of generating a spike at each time-step equals the input $I$. It can be proved that the resulting process is a homogenous (probability of spike generation constant over time-steps) Poisson process where the average firing rate, i.e. average number of spikes generated over the entire train duration, is given by \cite{heeger2000poisson},
\begin{equation}
\label{eq2}
<\widetilde{I}(t)>=\frac{\sum_{t} \widetilde{I}(t)}{T_{N}}=I
\end{equation}
The spiking neuron processes the input spikes and generates a set of output spikes $\widetilde{y}(t)$. The response of the neuron is determined by its average firing activity over the $T_{N}$ time-steps, $<\widetilde{y}(t)>$. Note that such input encoding and neuron output measurement schemes are standard norms for SNNs and is not an additional requirement/overhead for our proposal. Our proposal concerns the manner in which the neuron will process and generate the output spike train $\widetilde{y}(t)$. In order to achieve near lossless (with respect to accuracy) conversion from ANN to SNN, $<\widetilde{y}(t)>$ should approximate $y$ reasonably well.

Prior work on such ANN-SNN conversion replace the original ANN neural unit with an SNN unit that has little correspondence to the original ANN unit being considered and often rely on heuristic mechanisms to achieve the conversion, thereby incurring some accuracy loss in the process. Our conversion mechanism follows from the very intuitive observation that the analog activation output of the ANN neuron in the range $[0,1]$ can be mapped to the probability of spike generation, $p(t)$, of the spiking neuron at each time-step. Hence, at each time-step $t$, the neuron receives the input spike train, $\widetilde{I}(t)$, and generates an output spike with probability $p(t)=f(\widetilde{I}(t))$. 

Now, let us provide a mathematical analysis to justify that such a mapping is able to approximate the original ANN neural unit to a reasonable degree of precision. It follows from Eq. \ref{eq2}, that the spike train consists of $I.T_{N}$ number of spiking events and $(1-I).T_{N}$ number of non-spiking events, on the average, over the entire duration of time-steps, $T_{N}$. The output spike train is generated according to an inhomogeneous Poisson process \cite{heeger2000poisson} (spike generation probability varies over time), where the probability of spike generation is equal to $p(t \vert \widetilde{I}(t) = 1)=\frac{1}{1+e^{-w}}$ whenever there is an input spike and $p(t \vert \widetilde{I}(t) = 0)=\frac{1}{1+e^{0}}=\frac{1}{2}$ in the case of no spike. Hence, the inhomogeneous Poisson process can be decomposed into two homogeneous Poisson processes corresponding to spiking (of duration $I.T_{N}$ time-steps) and non-spiking events (of duration $(1-I).T_{N}$ time-steps). Hence, the average firing activity of the neuron will be given by the sum of the firing activities of the individual Poisson processes averaged over the total number of time-steps, $T_{N}$. Following Eq. \ref{eq2}, we can state that the average firing rate of the output spike train, $\widetilde{y}(t)$, is given by,
\begin{equation}
\label{eq3}
\begin{aligned}
<\widetilde{y}(t)>&=p(t \vert \widetilde{I}(t) = 1).I+p(t \vert \widetilde{I}(t) = 0).(1-I)\\
&=\frac{I}{1+e^{-w}}+\frac{1-I}{1+e^{0}}\\
&=\frac{1}{2}+\frac{I}{2}\left(\frac{1-e^{-w}}{1+e^{-w}}\right)
\end{aligned}
\end{equation}
Closer inspection of the above equation reveals that $<\widetilde{y}(t)>$ is a linear approximation of the sigmoid function in the range $I\in[0,1]$. Fig. \ref{fig1}(b) and (c) represents a plot of the outputs, $y$ (ANN) and $<\widetilde{y}(t)>$ (SNN) with variation in the input $I$ and for synaptic weight magnitudes $w=1$ and $w=3$ respectively (3 being the maximum weight for the synapses in our network). Note that the negative range for $I$ represents the case for negative synaptic weight. As can be concluded from the figure, the error between the functions is almost negligible for $w=1$ and increases slightly as the magnitude of the weight increases. However, even for the maximum weight $w=3$, the error remains bounded below reasonably low values over the entire approximation range. This fact is reinstated by Fig. \ref{fig1}(d) which represents a contour plot of the error magnitude between the two expressions $y$ and $<\widetilde{y}(t)>$ with variation in both $I$ and $w$. Note that since we are trying to encode information in the analog sigmoid output of the neural units, weights obtained as a result of backpropagation training typically remain bounded below values that ensure that the neuron outputs do not fall in the saturation regime of the sigmoid function. As can be observed from Fig. \ref{fig1}(c), for a weight magnitude of 3, almost the entire range of the sigmoid function is being used and hence it is expected that synaptic weights should converge to such limited ranges after the training process. Additionally neural nets, being inspired from computational mechanisms observed in the biological brain, are characterized by an inherent tolerance to variations in the neural and synaptic units and hence such minor variation between $y$ (ANN) and $<\widetilde{y}(t)>$ (SNN) is not expected to impact the network performance.

\section{MTJ as a Probabilistic Neuron}
The MTJ consists of a thin spacer layer, typically $MgO$, sandwiched between two nanomagnetic layers. While the magnetization state of one of the layers (``Free Layer'': FL) is switched from one stable state to another, the magnetization of the other nanomagnet (``Pinned Layer'': PL) is fixed. The relative orientation of the FL magnetization with respect to the PL magnetization determines the resistance of the MTJ. The MTJ exhibits a low resistive ``Parallel'' (P) state when both FL and PL magnetizations are in the same direction and a high resistive ``Anti-Parallel'' state (AP) otherwise. In this work, we will consider FL switching induced by spin-orbit torque generated by a heavy metal (HM) underlayer due to the potential of achieving energy-efficient switching along with the possibility of interfacing such a device as a neuron with a synaptic resistive crossbar array. These advantages will be discussed in details in the subsequent text. The underlying device phenomena that lends a randomness or probabilistic feature to the switching event of the MTJ is the inherent time-varying thermal noise. As we will show later, the probability of MTJ switching increases in a non-linear fashion as the magnitude of input current through the HM is increased.
\begin{figure}[!t]
\centering
\includegraphics[width=2.6in]{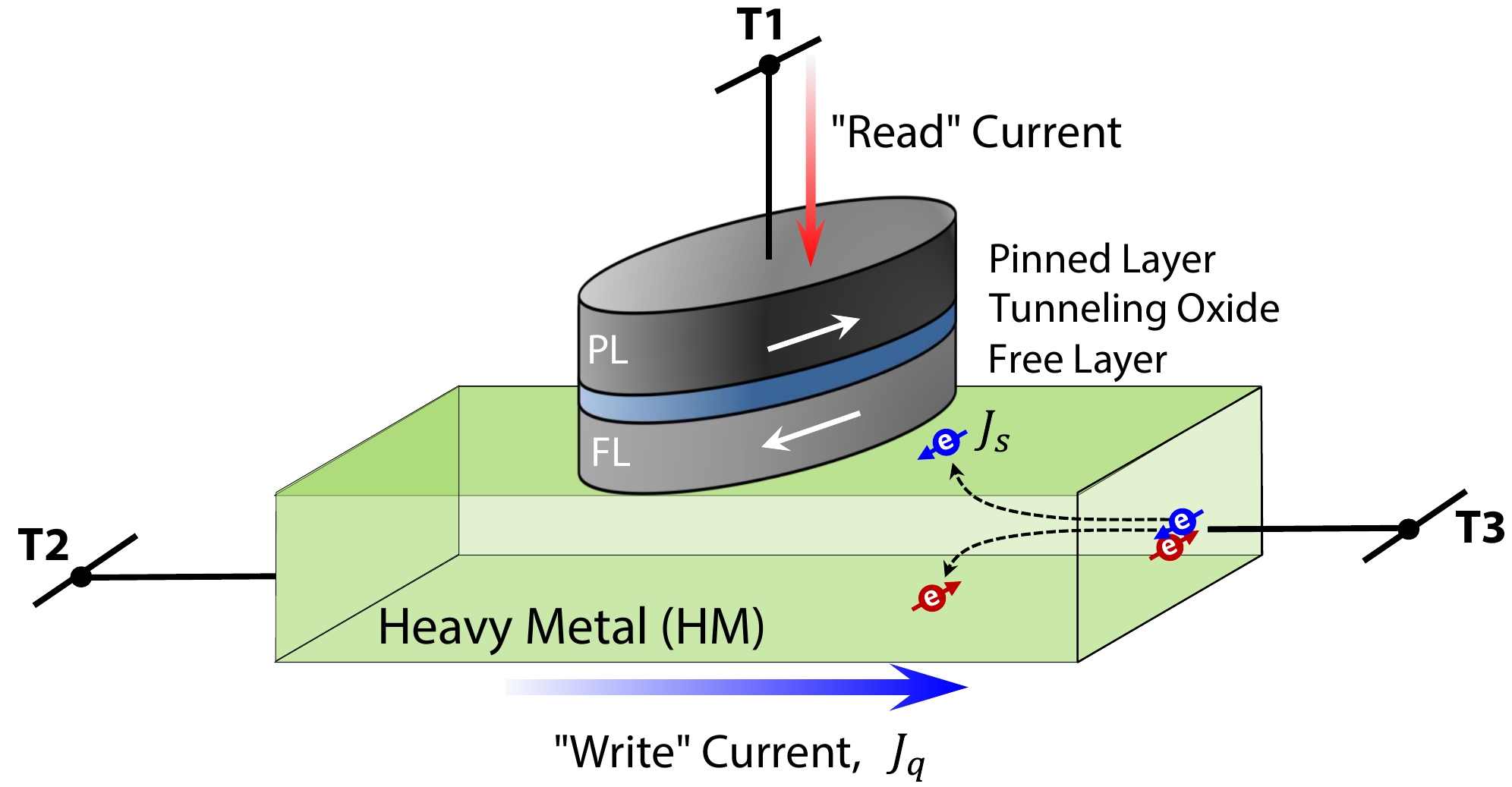}
\caption{Three terminal device structure consisting of an MTJ lying on top of a HM as a probabilistic spiking neuron. ``Write'' current flowing through the HM stochastically switches the MTJ in presence of thermal noise.}
\label{fig2}
\end{figure}

Let us first illustrate the underlying physical phenomena involved in spin-orbit torque (SOT) induced MTJ switching. The three terminal device structure under consideration is shown in Fig. \ref{fig2}. When an input charge current density, $J_q$, flows through the HM underlayer from terminal T2 to T3, an input spin current, $J_s$, is injected into the FL whose spin orientation is perpendicular to both $J_q$ and $J_s$ (assuming spin-Hall effect \cite{hirsch1999spin} to be the dominant mechanism). Hence, the injected spin current being in-plane polarized can be utilized to switch a FL with in-plane magnetic anisotropy. Note that such SOT induced MTJ switching has been confirmed by multiple experiments \cite{liu2012spin,pai2012spin,liu2012current,suzuki2011current} and the simulation framework considered in this work are based on measurements reported in \cite{pai2012spin}.

The input spin current density is related to the charge current density flowing through the HM underlayer by the relationship, $J_s = \theta_{SH}.J_q \implies I_s = \theta_{SH}.\left(\frac{A_{MTJ}}{A_{HM}}\right)I_q$, where $I_{s}$ and $I_{q}$ are the input spin current and charge current magnitudes respectively, $\theta_{SH}$ is the spin-Hall angle \cite{hirsch1999spin,pai2012spin} and, $A_{MTJ}$ and $A_{HM}$ are the MTJ and HM cross-sectional areas respectively. Although $\theta_{SH}<1$, the input spin current polarization can be maintained greater than $100\%$ ($J_s>J_q$) by properly optimizing the device dimensions ($A_{MTJ}$ and $A_{HM}$). In contrast, conventional spin-transfer torque (STT) switching due to charge current flow through PL is always limited by the spin-polarization strength of the PL ($<100\%$). In addition to the possibilities of achieving energy-efficient switching in comparison to conventional STT mechanism, the input current flows through the HM underlayer, which typically has a resistance of a few hundred ohms. This makes such a three terminal device a potential candidate to be operated as a neuron interfaced with a resistive crossbar array of synapses. Proper functioning of the array requires the input resistance of the neuron to be sufficiently low in comparison to the synaptic resistances at each cross-point (discussed in details in later text). In contrast, standard two-terminal MTJ with STT induced switching (charge current flowing from terminals T1 to T3 or T3 to T1 through the PL to generate the necessary spin current) would require the input current to flow through the oxide layer, which would typically have considerably higher resistance, thereby leading to non-ideal network operation. 

The probabilistic switching characteristics of the MTJ can be analyzed by Landau-Lifshitz-Gilbert (LLG) equation with additional term to account for SOT generated by the HM underlayer \cite{slonczewski1989conductance},
\begin{equation}
\label{llg}
\frac {d\widehat {\textbf {m}}} {dt} = -\gamma(\widehat {\textbf {m}} \times \textbf {H}_{eff})+ \alpha (\widehat {\textbf {m}} \times \frac {d\widehat {\textbf {m}}} {dt})+\frac{1}{qN_{s}} (\widehat {\textbf {m}} \times \textbf {I}_s \times \widehat {\textbf {m}})
\end{equation}
where, $\widehat {\textbf {m}}$ is the unit vector of FL magnetization, $\gamma= \frac {2 \mu _B \mu_0} {\hbar}$ is the gyromagnetic ratio for electron, $\alpha$ is Gilbert\textquoteright s damping ratio, $\textbf{H}_{eff}$ is the effective magnetic field including the shape anisotropy field for elliptic disks, $N_s=\frac{M_{s}V}{\mu_B}$ is the number of spins in free layer of volume $V$ ($M_{s}$ is saturation magnetization and $\mu_{B}$ is Bohr magneton), and $\textbf{I}_{s}$ is the spin current generated by the HM underlayer. Thermal noise is included by an additional thermal field \cite{scholz2001micromagnetic}, $\textbf{H}_{thermal}=\sqrt{\frac{\alpha}{1+\alpha^{2}}\frac{2K_{B}T_{K}}{\gamma\mu_{0}M_{s}V\delta_{t}}}G_{0,1}$, where $G_{0,1}$ is a Gaussian distribution with zero mean and unit standard deviation, $K_{B}$ is Boltzmann constant, $T_{K}$ is the temperature and $\delta_{t}$ is the simulation time-step.

\begin{table}[h]
\label{table}
\center
\centerline{TABLE I. Device Simulation Parameters}
\vspace{2mm}
\begin{tabular}{c c}
\hline \hline
\bfseries Parameters & \bfseries Value\\
\hline
Free layer area & $\frac{\pi}{4} \times 100 \times 40 nm^2$\\
Free layer thickness & $ 1.2 nm$\\
Heavy-metal thickness, $t_{HM}$ & $ 2 nm$\\
Saturation magnetization, $M_{S}$ & 1000 $KA/m$ \cite{pai2012spin}\\
Spin-Hall angle, $\theta_{SH}$ & 0.3 \cite{pai2012spin} \\
Gilbert damping factor, $\alpha$ & 0.0122 \cite{pai2012spin} \\
Energy barrier, $E_{B}$ & 20 $K_{B}T$ \\
Resistivity of HM, $\rho_{HM}$ & 200 $\mu\Omega.cm$ \cite{pai2012spin}\\
Pulse width, $T_{w}$ & $ 0.2, 0.5, 1ns$ \\
Temperature, $T_{K}$ & $300K$ \\
Read voltage, $V_{read}$ & $1V$ \\
\hline \hline
\end{tabular}\\ 
\end{table}
\begin{figure}[!t]
\centering
\includegraphics[width=3.4in]{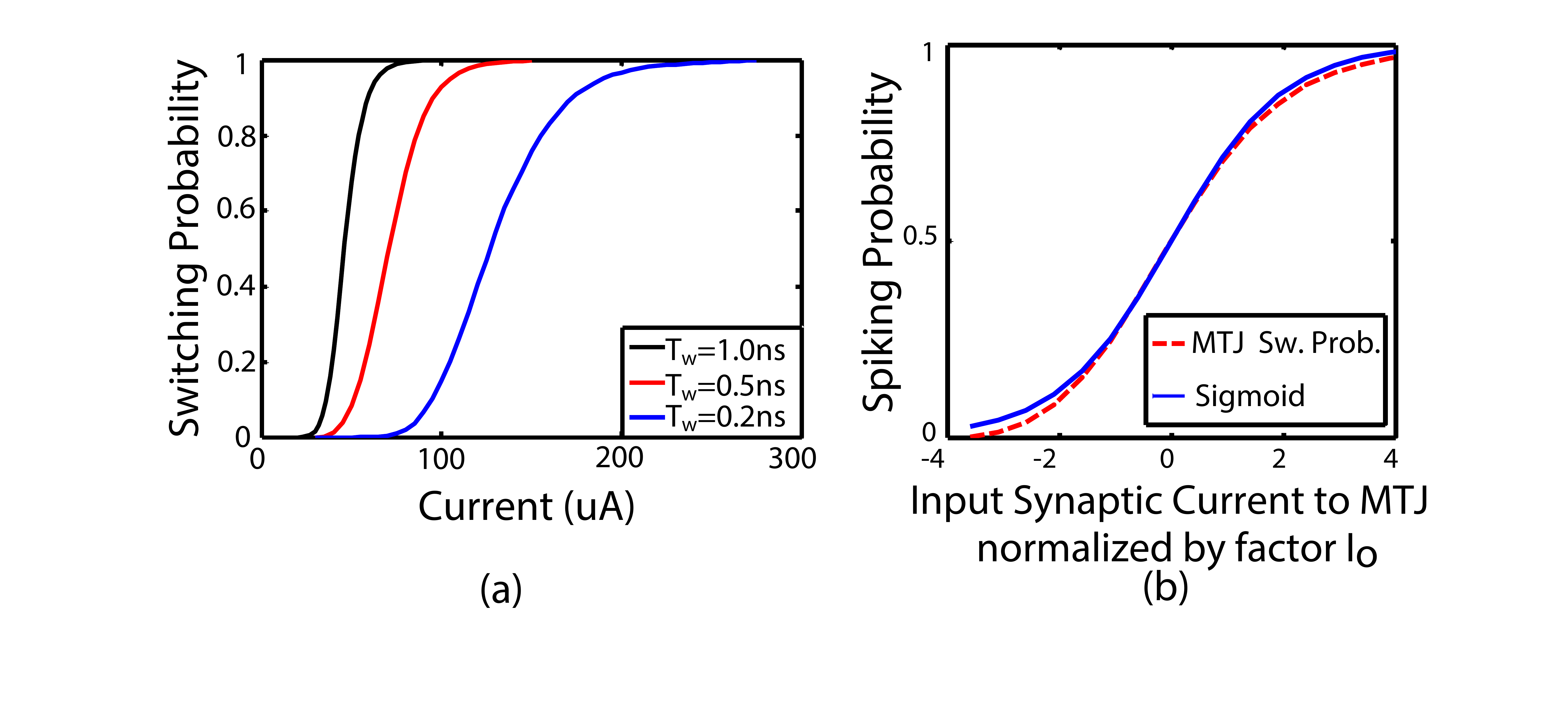}
\caption{(a) Switching probability characteristics of an MTJ of volume $\frac{\pi}{4} \times 100 \times 40 \times 1.2 nm^3$ at $T_{K}=300K$ for different ``write'' cycle durations, $T_{w}$, (b) MTJ switching probability characteristics re-plotted for $T_{w}=0.5ns$ as a function of the input synaptic current, $I_{syn}$, normalized by factor $I_{o}=10\mu A$. The data closely resembles a sigmoid probability density function.}
\label{fig3}
\end{figure}

The device simulation parameters have been outlined in Table. I and are based on experimental measurements performed in Ref. \cite{pai2012spin}. A barrier height of $20K_{B}T$ was chosen since the MTJ is being used as a computing element in this application. Fig. \ref{fig3}(a) depicts the switching probability of the MTJ with variation in the magnitude of input current. The probability switching characteristics undergoes more dispersion with decrease in the duration of the input ``write'' current, $T_{w}$. While more dispersion in the characteristics results in increased robustness of the system in presence of variations, power consumption of the network increases. These tradeoffs will be discussed in details in the next section. In order to map such switching probability characteristics of the MTJ to the sigmoid probability function for spike generation discussed in the previous section, the MTJ is considered to be driven by two input currents, namely $I_{bias}$ and $I_{syn}$. The current $I_{bias}$ provides the necessary current to the MTJ to bias it at a probability of 0.5. The current $I_{syn}$ is the resultant input synaptic current to the neuron. Hence, in absence of $I_{syn}$, the MTJ has $50\%$ probability of switching similar to the sigmoid characteristics. Fig. \ref{fig3}(b) illustrates the switching probability characteristics of the MTJ with variation in input synaptic current, $I_{syn}$ (normalized by a factor, $I_{o}$, which encodes the degree of dispersion of the MTJ switching probability characteristics). The switching characteristics match the sigmoid variation to a reasonable degree of approximation. Also, note that such neuromorphic algorithms are highly error-resilient and such small approximations in the neuron output will not cause significant changes in the network performance. We will validate our claims by presenting results for a large-scale deep neural network in the next section. The mapping of the normalization factor in the input synaptic current, $I_o$, to the hardware implementation of a synaptic crossbar array will be discussed later.
\begin{figure}[!t]
\centering
\includegraphics[width=3.4in]{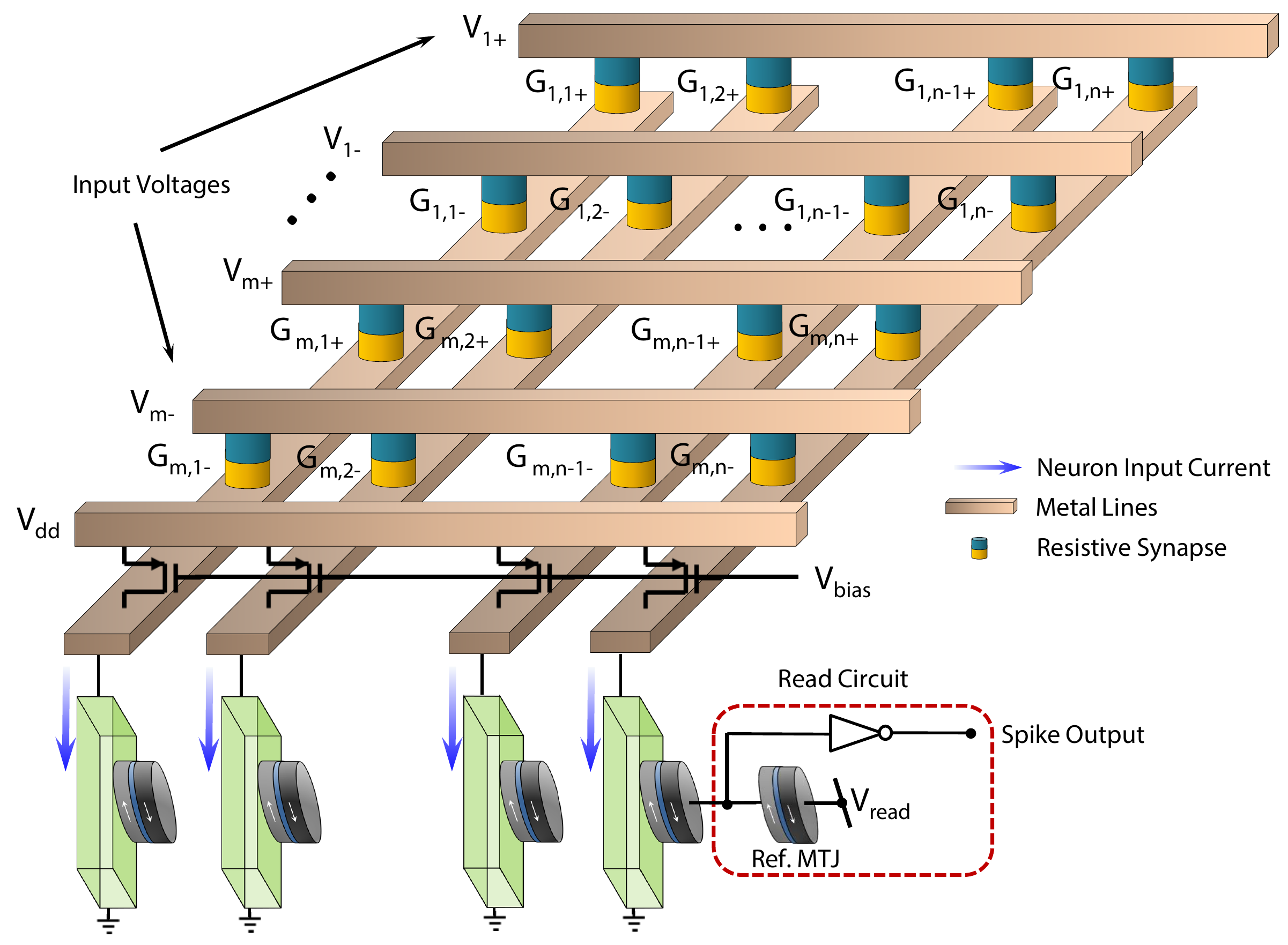}
\caption{Hardware mapping of the computing core (weighted synaptic summation of inputs followed by neural processing) mapped to a crossbar array of resistive synapses interfaced with probabilistic spiking MTJ neurons. The additional row of PMOS transistors supply the necessary input bias current to each of the MTJ neurons.}
\label{fig4}
\end{figure}

In order to implement a neural network, neurons need to be interfaced with synapses. The basic computing core in any neural network architecture, even for deep networks, consists of a dot product implementation where each of the neural inputs are initially multiplied by synaptic weights, and are subsequently processed by the neuron. Such a functionality can be directly mapped to a crossbar architecture, as shown in Fig. \ref{fig4}, where each horizontal metal line provides an input voltage across each resistive synaptic device and each vertical metal line provides an input current to the MTJ situated at the end of the vertical column. In order to implement bipolar weights, two rows ($V_{i+}$ and $V_{i-}$) are used for each input $V_i$. When the input $V_i$ assumes a logic value of \textquoteleft 0\textquoteright (no spike), then \textquoteleft 0\textquoteright \ voltage level is applied to both the inputs. However, when $V_i$ assumes a logic value of \textquoteleft 1\textquoteright (spike), then voltage $V_o$ is applied to the row corresponding to $V_{i+}$ and $-V_o$ is applied to the row corresponding to $V_{i-}$. If the weight $w_{i,j}$ for the $j$-th neuron and input $V_i$ is positive, then the conductance corresponding to $V_{i+}$ is programmed to $G_{i,j+}=w_{i,j}.G_{o}$ ($G_{o}$ is the mapped conductance for unity weight), while the conductance, $G_{i,j-}$ corresponding to $V_{i-}$ is programmed to high OFF resistive state and vice versa. It is worth noting here that the resistive synapses can be implemented by phase change devices \cite{jackson2013nanoscale,kuzum2011nanoelectronic}, memristive devices \cite{jo2010nanoscale} or even spintronic synapses \cite{sengupta2015spin}. Since synaptic learning is not the focus of this work (offline learning), a resistor model was considered for the synapses with 4 bit discretization in the synaptic levels and a maximum to minimum conductance ratio of 10, which are typical of such resistive memory technologies. 

Let us consider the conductance in the path of the net synaptic current while flowing through the HM of the spintronic neuron to be $G_s$ and the voltage drop across the neuron to be $V_s$. Equating the current supplied by the resistive synapses along with the input bias current, $I_{bias}$, to the current flowing through the neuron, we get $\sum\limits_{i}(G_{i,j+}.(V_{i+}-V_s)+G_{i,j-}.(V_{i-}-V_s))+I_{bias}=G_s.V_s$ which indicates that the net synaptic current supplied to the spintronic neuron is given by, 
\begin{equation}
\label{eq5}
\begin{aligned}
I_j&=\frac {G_s. \left(\sum\limits_{i}(G_{i,j+}.V_{i+}+G_{i,j-}.V_{i-})+I_{bias}\right)} {G_s+ \sum\limits_{i}(G_{i,j+}+G_{i,j-}))} \\
&=\frac{\sum\limits_{i}(G_{i,j+}.V_{i+}+G_{i,j-}.V_{i-})+I_{bias}}{1+\gamma}
\end{aligned}
\end{equation}
Note that the resultant weighted synaptic input (with respect to the computational model described in Section II) is scaled by a factor $G_{o}.V_{o}$ (in the current domain). Hence, in order to map the functionality to the sigmoid probability characteristics, the scaling factor in the MTJ switching characteristics discussed previously, $I_{o}$ has to be equal to $G_{o}.V_{o}$. In other words, the resultant synaptic current being supplied by the crossbar array needs to be adjusted according to the dispersion of the switching probability characteristics of the MTJ in order to maintain consistency with the computational model described previously. 

Another interesting point to note is the non-ideality factor, $\gamma =\sum\limits_{i}(G_{i,j+}+G_{i,j-})/G_{s}$ in Eq. \ref{eq5}. This reiterates the fact that the input resistance of the neuronal device has to be sufficiently low in order to ensure that most of the input voltage drops across the resistive synapses and the voltage drop across the neurons are negligible. Hence, a sufficient value of the spike voltage, $V_{o}$ (which dictates the value of $G_{o}$), has to be maintained to ensure that $\gamma<<1$. Duration of the input ``write'' current also has an impact on the choice of $V_{o}$ and $G_{o}$. With more duration of input current and hence, less dispersion in the switching characteristics, $I_{o}$ decreases resulting in decrease of $G_{o}$ and hence $\gamma$. However, robustness of the system to variations in the bias current and synaptic conductances suffer. These design space explorations will be considered in details in the next section.
Operation of each time-step of the SNN takes place through three cycles. In the first phase or the ``write'' cycle, the MTJ neuron receives the bias current and the input synaptic current from the crossbar array and switches probabilistically. Note that the bias current can be provided by an additional row of the crossbar array consisting of PMOS transistors biased in saturation. After the ``write'' cycle, the ``read'' terminals of the neuron are activated. As shown in Fig. \ref{fig4}, the ``read'' circuit consists of a resistive divider network with a ``Reference'' MTJ (whose state is fixed to the AP state). Hence a spike (logic value `1') is generated at the output inverter in case the MTJ switches to the P state. In case a spike is generated, the MTJ is switched back to the AP state by passing a sufficiently high magnitude of current through the HM in the opposite direction during a subsequent ``reset'' phase to ensure normal MTJ operation during the next time-step.
\section{Results and Discussions}
\subsection{Device-Circuit-Algorithm Co-Simulation Framework}
In order to validate the proposal and explore the design space of the network, a hybrid device-circuit-algorithm co-simulation framework was devised. The probabilistic magnetization switching characteristics of the MTJ was determined by running stochastic LLG simulations for different input current magnitudes and ``write'' cycle durations (Fig. \ref{fig3}). This behavioral model of the MTJ switching characteristics was utilized for subsequent system level simulations of the network. SPICE simulations (including Verilog-A model of the MTJ resistance \cite{fong2011knack}) were performed to assess the power and energy consumption of the crossbar array and the MTJ ``read'' circuit.

The performance of the network was assessed for a large scale deep learning network architecture \cite{IMM2012-06284} (28x28-6c5-2s-12c5-2s-10o) on a standard digit recognition problem based on the MNIST dataset \cite{lecun1998gradient}. The network consists of alternate layers of convolutional and subsampling operations. While the convolutional layers constitute the major computationally expensive component of the network and can be mapped to the synaptic crossbar array interfaced with MTJ neurons as described in the previous section, the subsampling layer simply performs an averaging operation over the spikes generated by the MTJ neurons over non-overlapping windows of the convolution output maps. The dimensions of the input MNIST images are 28x28, which are applied as input to the convolutional layer consisting of 6 convolutional kernels of size 5x5. The subsampling kernel is of size 2x2, and is followed by another convolutional layer comprising of 12 output maps, which in turn, is followed by another subsampling layer. The final layer consists of 10 neurons, each of which represents one of the ten digit classes. The network is trained using 60,000 training samples based on the methodology outlined in Ref. \cite{IMM2012-06284} \footnote{The related code can be found at \url{https://github.com/rasmusbergpalm/DeepLearnToolbox} (MATLAB Deep Learning Toolbox).}. Once the training is accomplished, the learnt weights are mapped to the synaptic conductances using the scheme mentioned in the previous section. All recognition accuracies mentioned in this text are with respect to the 10,000 test samples in the dataset. The baseline ANN network was trained with an accuracy of $98.56\%$ over the testing set. During the operation of the converted SNN, the image pixels are converted to Poisson spike trains where the average number of spikes generated over a given time window encode the corresponding pixel intensity. 

Note that a deep learning architecture is being used in this work since it has achieved high recognition accuracies in a large number of complex datasets. Further the architecture only dictates the manner in which the neurons and synapses are connected to form the network. However, our proposal holds true for any neural network topology since the basic computational elements and their mapping to crossbar architectures remain equally valid. We would also like to point out that improved training algorithms/network architectures to enhance the performance of the network in terms of recognition accuracy can be performed. However, the goal of this work is to demonstrate the applicability of the MTJ as a probabilistic spiking neuron that can potentially enable near-lossless (with respect to classification accuracy), low-power, low latency SNNs converted from trained ANNs. 
\subsection{Impact of MTJ ``Write'' Cycle Duration and Crossbar Supply Voltage on Network Performance}
Let us first describe the impact of ``write'' cycle duration on the performance of the network. With increase in the duration of the ``write'' cycle, the switching probability characteristics become sharper. Hence the synaptic current requirement from the crossbar array reduces. Further, the bias current magnitude also reduces since spin-orbit torque is exerted on the magnet for a longer duration of time. Hence, power consumption of the network is expected to reduce with increase in the magnitude of the ``write'' cycle duration. However, this occurs at the expense of delay since the network has to be operated over a number of time-steps and each time-step duration is directly related to the duration of the ``write'' cycle.
\begin{figure}[!t]
\centering
\includegraphics[width=3.2in]{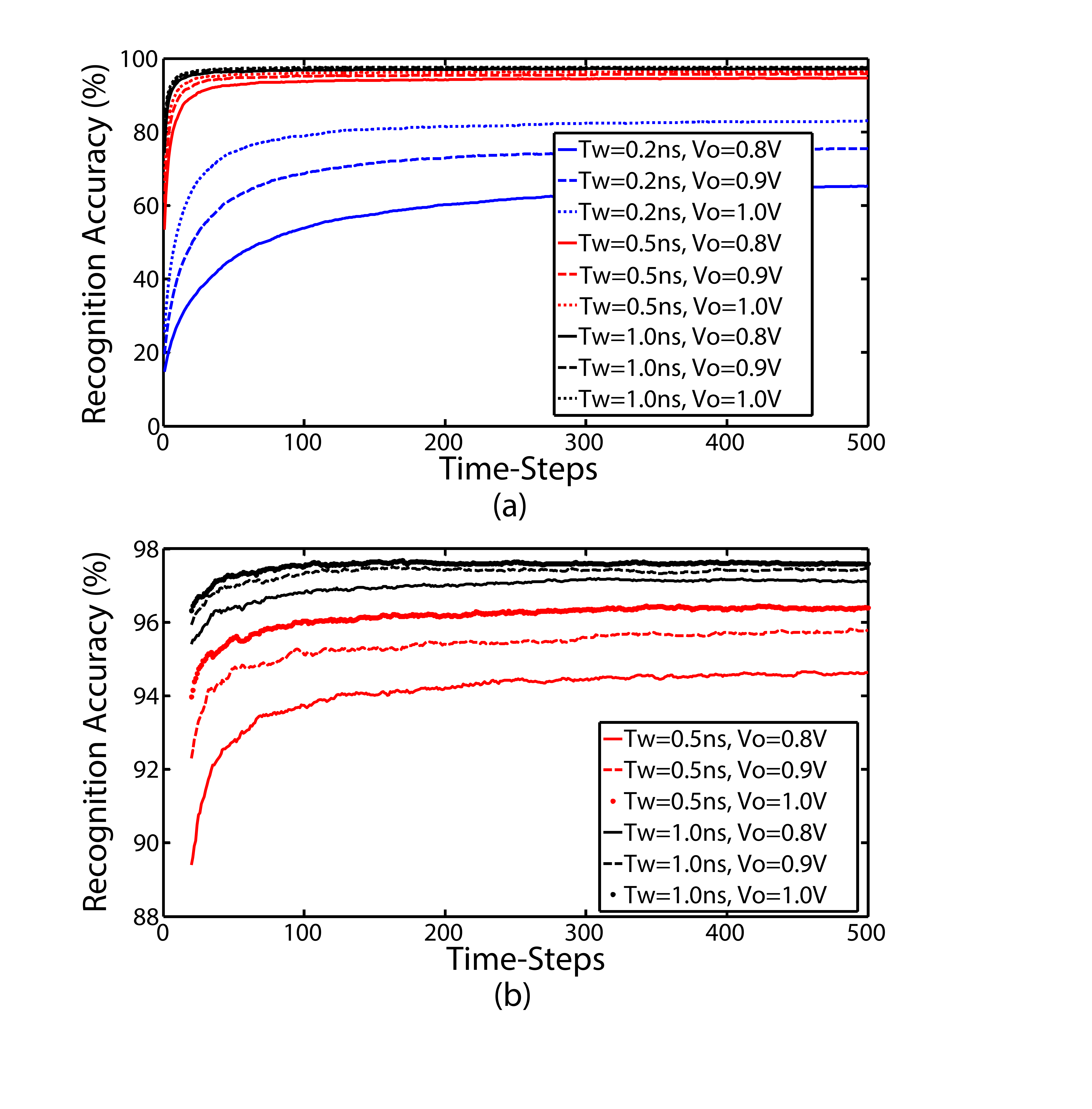}
\caption{(a) Recognition accuracy as a function of time-steps with variation in the ``write'' cycle duration ($T_w=0.2, 0.5$ and $1ns$) and crossbar supply voltage ($V_o=0.8, 0.9$ and $1V$), (b) Zoomed-in depiction of plot (a) from 50-500 time-steps for $T_{w}=0.5$ and $1ns$. Near-lossless SNN conversion can be achieved by maintaining a sufficient duration of the ``write'' cycle, even with scaling of crossbar supply voltage.}
\label{Result1}
\end{figure}

However, decrease in the ``write'' cycle duration, i.e. increase in the dispersion of the probability switching characteristics of the MTJ will result in increase of the factor $\gamma$, as discussed previously, thereby leading to non-ideal network operation. Fig. \ref{Result1} depicts the classification accuracy as a function of the time-steps of simulation of the SNN with varying ``write'' cycle durations ($T_{w}$), namely 0.2, 0.5 and 1$ns$. As expected, for a fixed supply voltage, classification accuracy improves with increase in the ``write'' cycle duration. While the network accuracy reaches $97.6\%$ and $96.4\%$ for $T_w=1ns$ and $0.5ns$ respectively, it saturates at $83\%$ for $T_{w}=0.2ns$ at the ends of 500 time-steps. An interesting point to note is the low latency in the performance of the network. The accuracy reaches $96.3\%$ and $93.8\%$ at the end of just 20 time-steps for $T_w=1ns$ and $0.5ns$ respectively. This is a crucial advantage offered by our ANN-SNN conversion scheme since although SNN implementations are ideal for low-power neural network implementations, they incur penalty in terms of the delay since the network outputs need to be observed over a number of time-steps to generate sufficient confidence in the inference process. With our proposed conversion scheme, network accuracies close to the original trained ANN baseline can be achieved only within a few tens of time-steps of the spiking network operation.

Scaling the supply voltage, in turn, results in increment of the factor $\gamma$, thereby leading to more errors in the network performance. However, it is worth noting here that the drop in recognition accuracy is minimal for sufficiently large durations of the ``write'' cycle. For instance, the accuracy drop is insignificant ($97.1\%$ and $94.6\%$ for $T_w=1ns$ and $0.5ns$ respectively) even with the crossbar supply voltage being scaled down to $0.8V$. The key point we would like to stress from this section is that by maintaining a sufficient duration of the ``write'' cycle, it is possible to achieve near-lossless SNN operation with minimal delay coupled with the possibilities of voltage scaling for reduction in power consumption. It is also worth noting here that the analysis performed in this section includes non-idealities arising from hardware mapping of the SNN to a synaptic resistive crossbar array interfaced with MTJ neurons (including non-ideality factor $\gamma$ and deviations of MTJ switching probability characteristics from ideal sigmoid function).
\subsection{Variation Analysis}
\begin{figure}[!t]
\centering
\includegraphics[width=3.4in]{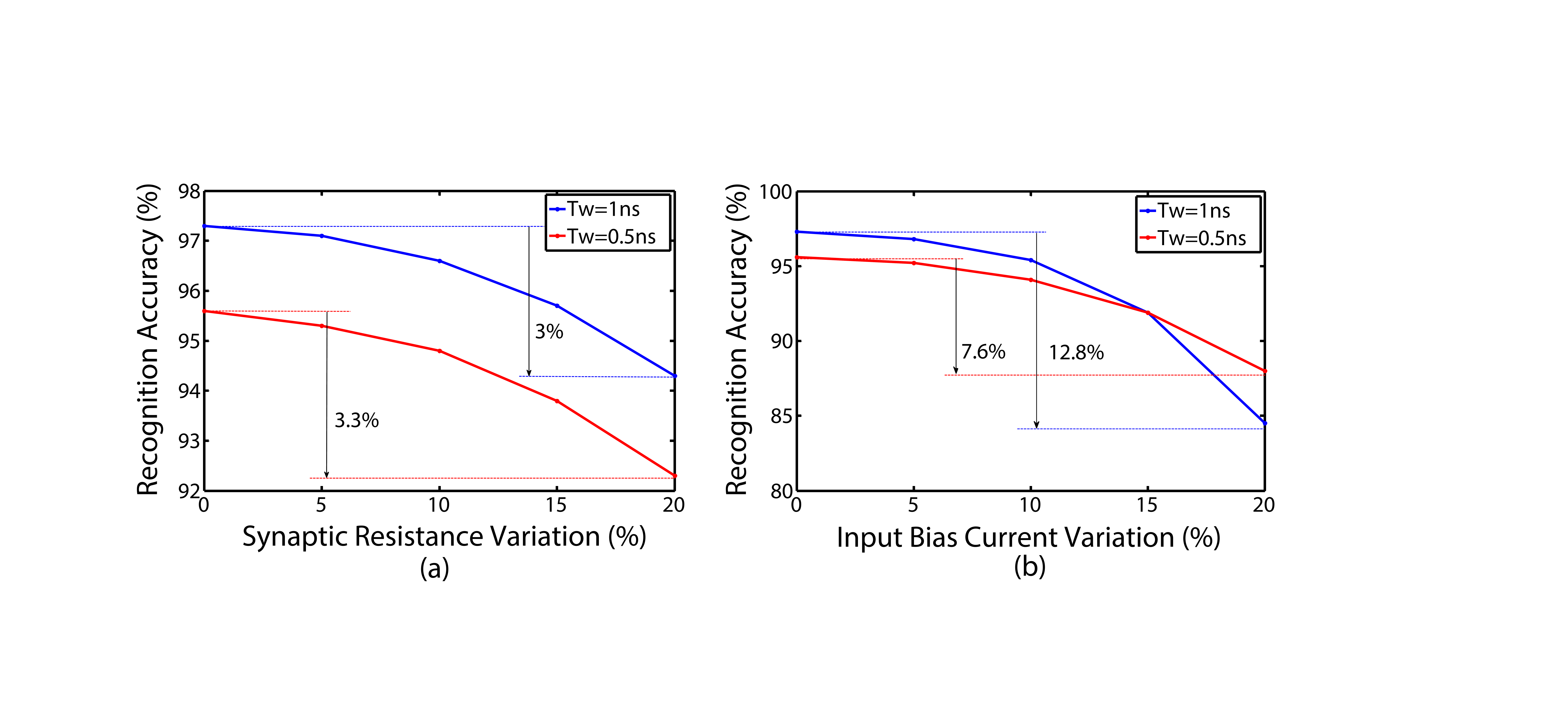}
\caption{Average recognition accuracy (measured over 50 independent Monte Carlo simulations for each of the 10,000 test images in the dataset) with variations (expressed as $\%$ $\sigma$ variation) in (a) resistances in the synaptic crossbar array and, (b) input bias current to the MTJ. The results have been measured at the end of 50 time-steps of SNN operation for crossbar supply voltage, $V_{o}=1V$.}
\label{Result2}
\end{figure}
Although increase in the ``write'' cycle duration helps to reduce the non-ideality in the network (by reduction of factor $\gamma$), it is associated with increased performance loss in presence of random variations due to sharper probability switching characteristics of the MTJ. In this section we will investigate the impact of random variations in the synaptic resistances of the crossbar array along with variations in the input bias current of the MTJ (Fig. \ref{Result2}). The average classification accuracy was determined by performing 50 independent Monte Carlo simulations of the network for each of the 10,000 test images in the dataset.

Fig. \ref{Result2}(a) depicts the average classification accuracy of the network with variations in the synaptic resistances of the crossbar array. Since the range of synaptic resistances are adjusted according to the dispersion of the MTJ switching probability characteristics (through the relation $I_o=V_o.G_o$ discussed previously), the impact of synaptic resistance variation is expected to be similar for different ``write'' cycle durations. An additional point to note is that, even with $\sigma=20\%$ variation in the synaptic resistances, only $3\%$ ($T_w=1ns$) and $3.3\%$ ($T_w=0.5ns$) degradation in classification accuracy was observed with respect to the original network (without variations) at the end of 50 time-steps. Such robustness to variations in the input synaptic current can be attributed to the error-resiliency of such neuromorphic computing systems.

However, the input bias current of the MTJ is a more critical parameter (with respect to variations) that ensures proper functionality of the network. Variations in the input bias current can skew the probabilistic MTJ operation in one direction, thereby causing degradation in recognition accuracy. Hence, sharper MTJ probability switching characteristics would result in more errors during the recognition process with variations in the input bias current. Fig. \ref{Result2}(b) illustrates that while $12.8\%$ reduction in accuracy was observed for $\sigma=20\%$ over the ideal network at the end of 50 time-steps for $T_w=1ns$, only $7.6\%$ degradation was observed for $T_w=0.5ns$. These results signify the fact that it is crucial to choose an optimal value of ``write'' cycle duration that simultaneously achieves near-lossless SNN conversion along with robustness to random variations in the input bias and synaptic currents. Note that a precise value of input bias current can be maintained by utilizing CMOS reference current generators that would exhibit $\sigma$ variations much less than $20\%$. However, impact on network performance with such high degree of variations was performed to establish that the network is highly error-resilient along with the fact that a judicious choice of the ``write'' cycle duration can enable robustness of the network even to large variations in the more sensitive MTJ input bias current. 

Additionally, we considered the impact of variation in the chip operating temperature by running a worst-case simulation where all the MTJs in the network were assumed to operate at $400K$ instead of the design temperature, $300K$. A recognition accuracy of $96.73\%$ was achieved at the end of 50 time-steps of network operation, thereby confirming that the proposed probabilistic neural computing framework is resilient to temperature variations as well.
\subsection{Power and Energy Benefits}

In order to evaluate the energy consumption of the network, SPICE simulations were performed to determine the energy consumption involved in ``write'', ``read'' and ``reset'' operations. In addition to providing a compact implementation of a spiking neuron, the MTJ enables low-power operation of the synaptic crossbar array. This is due to the fact that only input current magnitudes of a few tens of $\mu A$ need to be supplied by the crossbar array on either side of the bias current (Fig. \ref{fig3}). Note that the dominant power consumption of the network is involved in the synaptic crossbar array (since the number of synapses typically outnumber the number of neurons in such deep neural networks by two to three orders of magnitude), and such magneto-metallic spintronic neurons enable the low-power operation of the crossbar architectures. For the energy analysis, we considered the optimal ``write'' and ``reset'' cycle duration to be $0.5ns$ due to the possibilities of achieving near-lossless SNN conversion along with robustness to input bias current variations. An intuitive insight to the power efficiency of the network can be obtained by considering the fact that only $71\mu A$ of input current is required to bias the MTJ at $50\%$ switching probability ($T_w=0.5ns$). This current flowing through a HM resistance of $400\Omega$, results in an $I^{2}Rt$ energy consumption of $\sim 1 fJ$ in the neuron. Considering the resultant energy consumption in the ``write'', ``read'' and ``reset'' cycles of the network over a duration of 50 time-steps (since competitive classification accuracy can be obtained at the end of a few tens of time-steps), the total energy consumption of the proposed MTJ based SNN network was evaluated to be $19.5nJ$ per image classification.

An interesting point to note is that there is an additional delay overhead involved in the SNN operation. On the other hand, ANN operation (for instance, resistive crossbar array driven by analog CMOS neurons) would require a single time-step for recognition. However, the delay overhead (few tens of time-steps) is much smaller than the corresponding reduction in power consumption due to event (spike)-driven hardware operation. For example, the average energy consumption of an analog CMOS neuron is estimated to be $\sim 700fJ$ \cite{sharad2013spin} which would still be an order of magnitude greater than the average energy consumption of an MTJ neuron ($\sim 1fJ$) operated over a duration of 50 time-steps.

In order to compare with a baseline digital CMOS implementation, a deep spiking network consisting of Integrate-Fire (IF) neurons converted from a corresponding trained ANN was used based on the methodology proposed in Ref. \cite{diehl2015fast} for the same network architecture (28x28-6c5-2s-12c5-2s-10o) being considered in this work. The network was synthesized using a standard cell library in $45nm$ commercial CMOS technology. The design consisted of digital adders to sum up the synaptic weights in case of a spiking event (enabled by multiplexers). A comparator was utilized to compare the accumulated synaptic contributions to a specific threshold (IF functionality) and determine the corresponding spiking activity. A pipelined design with power-gating (to exploit the advantage of event-driven operation of the network) was considered with the same bit-discretization in the synaptic weights as mentioned previously. The average energy consumption involved in the network per image classification was evaluated to be $391nJ$ ($20\times$ more energy consumption than the proposed MTJ based spiking architecture).
\section{Summary}
In conclusion, we proposed a probabilistic neural computing platform that exploits the stochastic device physics of the MTJ to model complex neural transfer functions in the time domain. While the stochasticity of MTJ switching has been traditionally viewed as a disadvantage for logic and memory applications, we demonstrated that such probabilistic switching behavior can not only lead to high-accuracy cognitive recognition platforms but also provide energy benefits over conventional CMOS designs.

\end{document}